# Antibacterial Activity of Zinc Oxide Thin Films by Atomic Layer Deposition for Personal Protective Equipment Applications

Li Tao, Kinga Vojnits, Man In Lam, Xuejun Lu, Sepideh Pakpour, [*] Jian Liu [a, *]

*School of Engineering, Faculty of Applied Science, The University of British Columbia, 3333 University Way, Kelowna, BC V1V 1V7, Canada*

Corresponding authors' emails: sepideh.pakpour@ubc.ca (S. Pakpour), jian.liu@ubc.ca (J. Liu)

**Abstract**

The global pandemic has significantly increased the demand for personal protective equipment (PPE). The antimicrobial coating has been broadly applied to PPE to improve its prevention capability, especially after prolonged usage. However, antimicrobial coating by traditional methods, such as chemical vapor deposition, spraying, and slurry coating, suffers from drawbacks such as low efficiency, poor coverage, and loose adhesion to PPE. To overcome these limitations, this work adopted an atomic layer deposition (ALD) technique to deposit a zinc oxide (ZnO) thin film (~ 312.8 nm thick) as an antimicrobial coating and proved its advantage for depositing uniform ZnO coating on PPE with a fabric structure. Analysis by X-ray diffraction, Raman spectroscopy, and X-ray photoelectron spectroscopy confirmed the crystal structure and chemical composition of the ALD-ZnO. Ultraviolet-visible spectra disclosed a high absorption level of about 4.8 from 200 nm to 380 nm wavelength for ALD-ZnO, in contrast to 3.8 for commercial ZnO powders. Moreover, the ALD-ZnO exhibited strong antimicrobial properties when tested against *Escherichia coli (E.Coli)* in contrast to the control and bare glass samples. . The colony-forming unit (CFU/ml) remained zero for all ALD-ZnO samples while varying between $2.30\times10^9$ and $4.97\times10^9$ with a median of $4.36\times10^9$ for the control and between $9.10\times10^8$ and $3.27\times10^9$ with a median of $2.04\times 10^9$ for the bare glass. Statistic analysis using null-hypothesis significance testing revealed that the calculated *P* value, between bare glass and control, ALD-ZnO and control, and bare glass and control, were all smaller than 0.0001 and significantly smaller than 0.05 alpha value, suggesting a high confidence level of ALD-ZnO as the main factor for preventing *E.Coli* growth.

**Keywords:** Atomic layer deposition, antimicrobial coating, zinc oxide, personal protective equipment, bacteria

**1. Introduction**

Contamination caused by viruses and bacteria is always considered a significant threat to the general public's health. This is a particularly crucial problem for areas like medical service, food processing, and public transportation as they pose a high risk of spreading harmful microorganisms [1]. As human health and disease control concepts started to gain more awareness among the public [2], especially with the current impact of Corona Virus Disease 19 (COVID-19), the demand for new antibacterial and disinfection techniques also dramatically increased. Traditional cleaning methods are usually conducted by wiping contaminated surfaces with liquid cleaners, which are generally time-consuming and only effective for a short period. Moreover, not only are some cleaners hard to be obtained by the general public, but they can also be toxic to people and the environment [3]. Improper usage of aggressive compounds can even increase the potential of creating resistant bacteria and viruses, which can decrease future disinfection efficiency [4].

Recent investigations have been focused on developing inexpensive and safe antimicrobial materials for PPE applications [5]. Among many newly developed antimicrobial materials, photocatalytic self-disinfecting metal oxides have gained significant attention due to their low toxicity, long life, and good performance. Most metal oxide materials, such as titanium dioxide ($TiO_2$) and zinc oxide (ZnO), are chemically and physically stable. They also have a relatively low toxic level. As such, leaching such material into the environment causes much less negative impact than many other materials [6, 7]. However, metal oxides, such as $TiO_2$ and ZnO, are typically found in solid powder form, making applications of such materials as a self-disinfecting agent more difficult than liquid chemicals such as ethanol. To ensure the functionality of these materials, they are often made into slurries or mixed with binding agents, then pasted, painted, or mixed over or within desired application surface [8-12]. The disadvantages of these applying methods are apparent. Mixing metal oxide powders with other materials will reduce contact between metal oxides and microorganisms, lowering general antimicrobial efficiency. Depending on the properties of the binding material, the antimicrobial coating may also be worn out over time, making the disinfecting performance much less effective. More importantly, the binding material will also likely block the microstructure of metal oxide powders, decreasing the overall performance [13-15]. Last but least, these methods are difficult to achive uniform coating and complete coverage on PPE with complex structure, such as fabric and resuable masks.

The atomic layer deposition (ALD) technique is introduced herein to overcome the abovementioned limitations and achieve uniform antimicrobial coating on substrates with a large surface area. Introduced by Suntola in 1977, ALD is a thin film deposition method based on the sequential use of self-limiting surface reactions [16] [17]. Compared with popular film deposition methods, such as Chemical Vapor Deposition and Physical Vapor Deposition methods, the ALD technique can produce ultra-thin and ultra-uniform films with superior surface coverage even over complex three-dimensional (3D) geometries. Moreover, ALD deposition is generally performed at low temperatures from 350 °C down to room temperature, making it suitable for depositing an antimicrobial coating on temperature-sensitive PPE. Furthermore, benefiting from the chemical adsorption reaction and formation of chemical bonds, the ALD-produced films exhibit good adhesion to the substrate surfaces. With all the advantages mentioned above, the ALD technique is a promising approach for depositing metal oxide films to various PPE substrates and providing them with antimicrobial properties.

Herein, we deposited ALD-ZnO thin film on glass at 150 °C using diethylzinc and water as precursors. The chemical and physical properties of the ALD-ZnO were confirmed using various characterization techniques. Moreover, strict antimicrobial property valuation was performed on the ALD-ZnO, control, and bare glass following AATCC-100 standard assay, with *Escherichia coli (E. Coli)* being the model target microorganism. Colony counting tests clearly showed that the ALD-ZnO could effectively suppress the growth of *E. Coli*, reaching a similar level as commercial ZnO powders. It is expected that the ALD-ZnO could be rapidly used as a universal methodology to apply the antimicrobial coating for various PPE to protect people's health.

## 2. Experimental

*2.1 Deposition of ZnO coating by ALD*

The deposition of ZnO coating was performed on 22 × 22 mm glass substrates (Thermo Scientific cover, 0.18~0.25 mm thickness) in a plasma-enhanced atomic layer deposition (PEALD) system (GEMStar XT-R, Arridance, USA). All glass substrates were cleaned with isopropyl alcohol, followed by a rinse of deionized water, and completely dried before use. A custom-made steel grill was used to place and hold glass pieces. Glass pieces were attached to the grill by taping a small area on the corner with heat-resistant tapes, and compressed air was used to blow off surface dust before the grill was put inside the ALD reaction chamber. The ALD process of ZnO employed diethylzinc (DEZ, ⩾96%, VWR) and deionized water ($H_2O$) as the metal precursor and oxidizer, respectively. The source temperature for both DEZ and $H_2O$ was 25°C. The ALD reaction chamber and the chamber door were heated to 150°C and maintained such temperature during the reaction.

The surface reactions for ZnO with DEZ and $H_2O$ can be described in two half-reactions below [18-20]:

$$|\text{-OH}^* + Zn(C_2H_5)_2(g) \rightarrow |\text{-O-Zn}(C_2H_5)^* + C_2H_6(g) \quad (1)$$

$$|\text{-O-Zn}(C_2H_5)^* + H_2O(g) \rightarrow |\text{-O-Zn-OH}^* + C_2H_6(g) \quad (2)$$

Where the symbol "|" represents the substrate surface, "*" denotes the surface species, and "(g)" standards for a gaseous phase.

The overall reaction is:

$$Zn(C_2H_5)_2(g) + H_2O(g) \rightarrow ZnO + 2C_2H_6(g) \quad (3)$$

One ALD cycle of ZnO consists of four steps (Figure SI-1):

Step 1: Pulse DEZ vapor for 0.5 seconds and hold it for 5 seconds in the chamber for a complete surface reaction (Equation 1):

Step 2: Purge the ALD chamber with an Ar gas (50 sccm) to remove reaction byproduct $C_2H_6$ and excessive DEZ;

Step 3: Pulse $H_2O$ vapor for 0.2 seconds and hold it for 5 seconds in the ALD chamber (Equation 2);

Step 4: Pulse the ALD chamber with an Ar gas (50 sccm) for 15 seconds to remove the reaction byproduct $C_2H_6$ and extra $H_2O$.

The ALD-ZnO coating on the glass substrate was produced by repeating the above ALD cycle 2000 times, and the obtained sample was designated as ALD-ZnO on glass.

*2.2 Structural characterizations*

The thickness of the ALD-ZnO on glass was measured using a surface profilometer. Five measurements were performed at five random points on the sample surface, indicating an average surface thickness of 312.8 nm with a standard deviation of 41.3 nm. The surface morphology and topography of the ALD-ZnO on glass were observed using Scanning Electron Microscopy (SEM, Tescan Mira 3 XMU) and Atomic Force Microscope (AFM, Bruker Dimension Icon), respectively. The sample's chemical composition was analyzed using X-ray Photoelectron Spectroscopy (Kratos Analytical Axis Ultra DLD). The crystal structure of ALD-ZnO on glass was further analyzed with X-ray diffraction (XRD, Bruker D8 Advanced Powder X-Ray Diffractometer, 10 - 80°) and Raman spectrum (Bruker SENTERRA II Raman Microscope Spectrometer). UV absorption and transmission properties were analyzed using an ultraviolet-visible (UV-vis) spectrophotometer (Shimadzu UV-2700i Dual Beam Spectrophotometer). Commercial ZnO powders (GRADE CR-4, Ever Zinc) were taken as the baseline to compare with ALD-ZnO on glass.

*2.3 Experiment environment and setup*

The microorganism-related operations (bacteria culturing, planting, and diluting) were performed in a Biosafety Cabinet (Thermo Scientific 1300 series A2). To eliminate potential contaminations, each piece of equipment was cleaned with 70% ethanol or autoclaved at 120 °C for 30 minutes before use. Light-irradiation and antimicrobial evaluation were performed by placing the prepared samples inside a custom-made light chamber (Figure SI-2). Glass panels of this chamber are Electrochromic (EC) smart windows glass containing electrochromic coatings that could adjust the light passing rate by the tint control switch. A xenon arc high-power lamp (Sciencetech) generated a light spectrum similar to natural sunlight, simulating solar irradiations. Actual light passing into the chamber was controlled by changing the tint levels of the glass. A program-controlled humidifier was used to manage the relative humidity level inside the chamber, and conditions were monitored by an environment monitoring sensor (HOBO MX1104 ). Throughout each test, the chamber environment was maintained at 25 °C, 40% relative humidity, and tint level 2 (UVA/B intensity of ~ 54 µW cm$^{-2}$) to ensure consistent growth of *Escherichia coli (E.Coli)*.

*2.4 Antimicrobial activity analysis*

AATCC-100 standard assay was chosen as the antimicrobial performance analysis technique. It provided a quantitative procedure for comparing and evaluating the degree of antibacterial activity after exposure to the test bacteria on the test material compared directly against an untreated control.

As shown in Figure 1, *E.Coli* (ATCC 11229) was chosen as the target bacteria. *E.coli* stock was purchased from the American Type Culture Collection (ATCC), and fresh bacteria inocula were prepared by mixing *E. coli* stock (20 µm) into the autoclaved (120 °C, 30 minutes) nutrient broth (5 ml, BD 234000) inside a standard sterilized centrifugal tube (10 ml). The centrifugal tube was placed into a shaker at 37 °C with a shaking frequency of 90~150 rpm for 12 hours. A centrifugal tube with only nutrient media was also placed under the same condition as the centrifugal tube with *E.coli* for contamination controls. Optical Density (OD) values were measured with the reference of standard autoclaved nutrient broth to calculate the estimated bacterial cell growth.

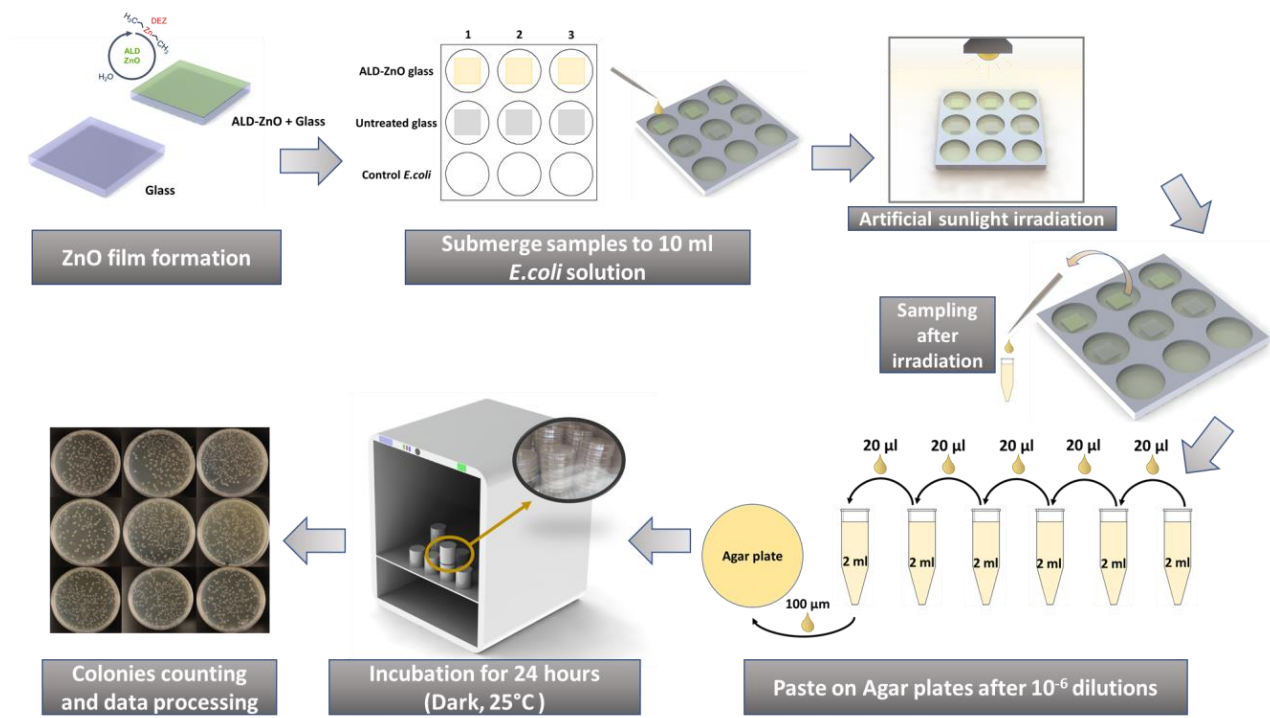

**Figure 1.** The procedure ALD-ZnO antimicrobial performance analysis.

ALD-ZnO on glass was first autoclaved at 120 °C for 30 minutes and placed inside a 50 mm diameter petri dish, submerged in 7 ml of *E. coli* nutrient broth solution with $1 \times 10^5$ cells/ml concentration. These Petri dishes were placed into a desired artificial growing environment with the optimized parameters (25 °C, 40% relative humidity, and UVA/B intensity of ~ 54 µW cm$^{-2}$) for 24 hours. Nutrient agar plates were prepared by pouring 20-ml autoclaved nutrient agar solution inside 100 mm diameter sterilized Petri dishes. Once the liquid nutrient agar solution solidified, Petri dishes were sealed and kept at 4 °C in a freezer for temporary storage.

After 12 hours of exposure, samples were removed from the light chamber, and serial dilutions were made with autoclaved nutrient broth solution. Dilution of $10^{-6}$ was subjected to colony counting tests. For this, 100 µm were spread onto the previously prepared and stored agar plates and then incubated in a dark incubator with temperature controlled at 37 ºC for 24 h. During this period, if the injected solution contains a sufficient amount of bacteria, bacteria colonies form on the surface of the agar as visible dots. After incubation, bacterial colonies were counted and recorded as colony-forming units (CFU/mL).

Antimicrobial activity, or bacteriostatic reduction (%), was calculated according to the equation below:

$$100(B-A)/B = BR \qquad (4)$$

Where *BR* is percentage of bacteriostatic reduction (%), *A* is the number of bacteria recovered from the inoculated treated test specimen incubated over the 24 h contact period, and *B* is he number of bacteria recovered from the inoculated untreated test specimen incubated over the 24 h contact period. Collected data was organized and processed with data analysis methods such as P-value test.

## 3. Results and Discussion

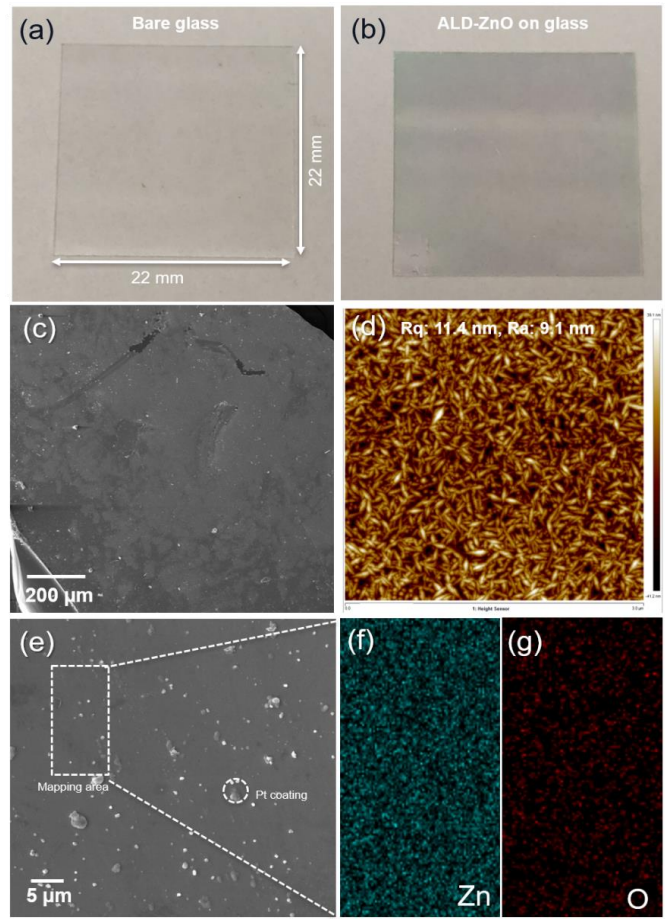

**Figure 2.** Visual comparison of (a) the bare glass and (b-g) ALD-ZnO on the glass; (c) SEM and (d) AFM images, and (e-g) EDX elemental mapping Zn and O elements of ALD-ZnO on the glass.

Figure 2 shows the morphology and topography of ALD-ZnO on the glass. Visual images in Figures 2a and 2b show that ALD-ZnO film on the glass presents a slight color change compared to the bare glass. SEM image in Figure 2c reveals good uniformity of ZnO on the glass, except for the Pt particles deposited by sputtering before SEM observation. A High-resolution AFM image (Figure 2d) indicates that the ALD-ZnO thin film consists of a large quantity of evenly distributed and densely packed nano-size plates. The RMS surface roughness ($Rq$) and average roughness ($Ra$) of ALD-ZnO is determined to be 11.4 and 9.1 nm, respectively. EDX mapping on the rectangle area in Figure 2e verifies the uniform distribution of Zn and O elements on the glass substrate (Figures 2f and 2g).

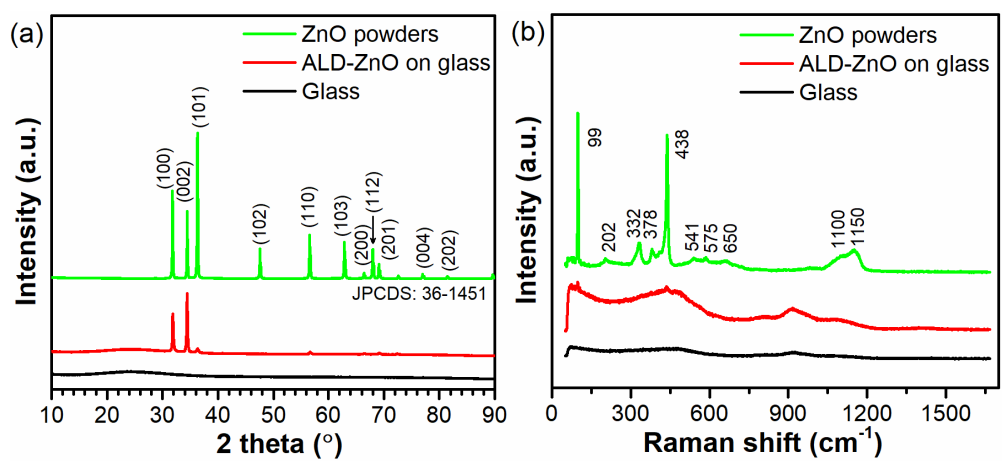

**Figure 3.** (a) XRD and (b) Raman spectra of ALD-ZnO on glass, commercial ZnO powders (GRADE CR-4, Ever Zinc), and bare glass substrates.

XRD and Raman spectroscopy was adopted to analyze the crystal structure of ALD-ZnO with ZnO powders and bare glass substrate as references, and the results are displayed in Figure 3. As shown in Figure 3a, the XRD pattern of ALD-ZnO clearly shows three peaks at 31.8°, 34.4°, and 36.3°, which can be indexed as the (100), (002), and (101) planes of ZnO (JPCDS PDF No. 36-1451), respectively. It should be noted that the relative peak intensity of (100), (002), and (101) changes for ALD-ZnO compared to commercial ZnO powders. This could be due to the different forms of the ZnO, *i.e.*, thin film for ALD-ZnO and powders for commercial ZnO. It is speculated that the (002) plane is the preferred growth direction for ZnO thin film deposited by ALD. Moreover, Figure 3a shows that the ALD-ZnO diffraction peaks are strong and shield the broad diffraction peak from the glass substrate. The Raman spectra in Figure 3b disclose relatively weak peaks at 99 and 438 cm$^{-1}$ for ALD-ZnO on glass, in sharp contrast with commercial ZnO powders. The broad background in the Raman spectrum of ALD-ZnO results from the glass substrate. Nevertheless, the results above confirm the successful deposition of uniform ZnO thin film on the glass substrate from the ALD process of DEZ/H$_2$O .

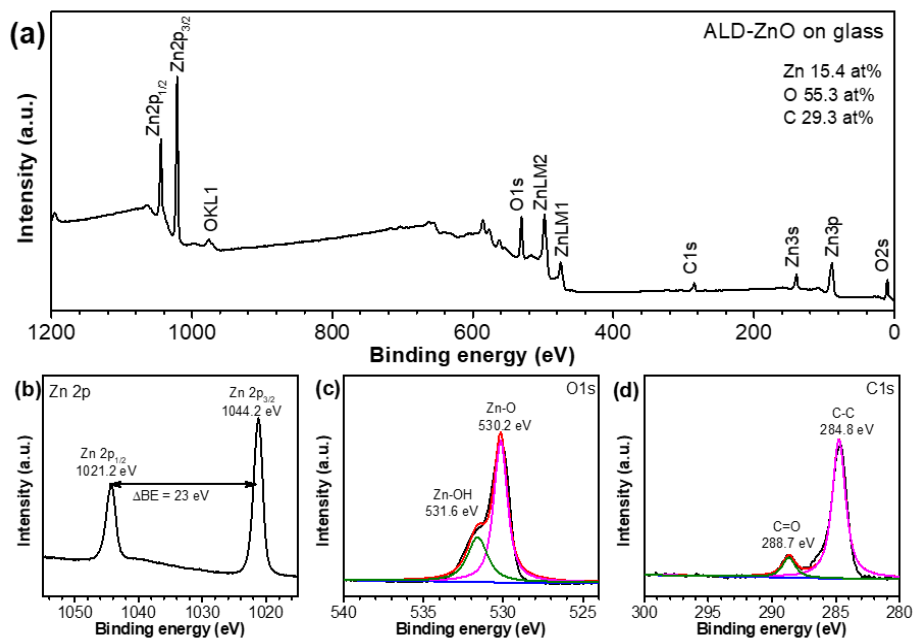

**Figure 4.** (a) XPS survey spectrum and high-resolution XPS spectra of (b) Zn 2p, (c) O1s, and (d) C1s for ALD-ZnO on the glass substrate.

XPS characterization was performed on the ALD-ZnO glass to confirm its chemical composition and bonding, and the results are displayed in Figure 4. XPS survey indicates the existence of Zn (15.4 at%), O (55.3 at%), and C (29.3%) in the ALD-ZnO thin film. A high-resolution scan of the Zn 2p spectrum (Figure 4b) results in two distinct peaks at 1021.2 and 1044.2 eV, corresponding to Zn $2p_{1/2}$ and Zn $2p_{3/2}$ of Zn-O with spin-orbit splitting energy of 23.0 eV ($\Delta BE$), respectively. This result agrees with the value of ZnO reported previously [21, 22]. Deconvolution of the O1s spectrum leads to two peaks at 530.2 eV and 531.6 eV (Figure 4c), which belong to the Zn-O bonding in ZnO and the -OH group on the ALD-ZnO surface, respectively [23]. A large amount of C impurity (29.3%) is detected in the ALD-ZnO thin film, and high-resolution XPS analysis reveals the coexistence of C=O (288.7 eV) and C-C (284.8 eV). The C impurity could result from residual unreacted precursors on the surface and adventurous carbon from the air.

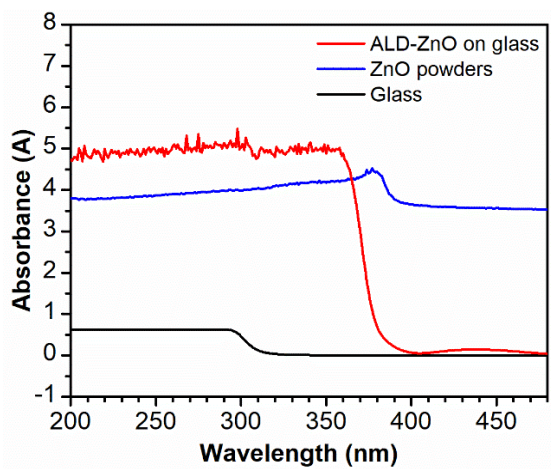

**Figure 5.** Absorbance spectra of ALD-ZnO on glass and ZnO powders analyzed by UV-Vis spectroscopy.

Photocatalytic materials have to interact with photons to establish photocatalytic processes, which would result in light absorption. Ultraviolet-visible spectroscopy (UV-vis) test was performed to measure the absorption of UV light on ALD-ZnO on glass and commercial ZnO powders at room temperature to investigate their photocatalytic behavior and light absorption properties, and the results are compared in Figure 5. The bare glass sample demonstrates no UV absorption until below 320 nm wavelength, with a steady but low absorption level at around 0.75. In contrast, the ALD-ZnO on glass displays a rapid growth of absorption level at about 380 nm wavelength and reaches an absorption level of about 4.8. The high absorption level spanned from 200 nm to 380 nm wavelength, covering most wavelengths of UV-A (315–400nm) and UV-B (295–315 nm) [24, 25]. The commercial ZnO powders exhibit a high absorption level at around 3.8 throughout the entire wavelength scan, probably due to the low transparency of the powder sample. However, it demonstrates a slight absorption increase at about 38 5nm as the wavelength enters the UV-A range. Based on the results above, it can be concluded that the ALD-ZnO coating has the highest average absorption level among all three samples tested.

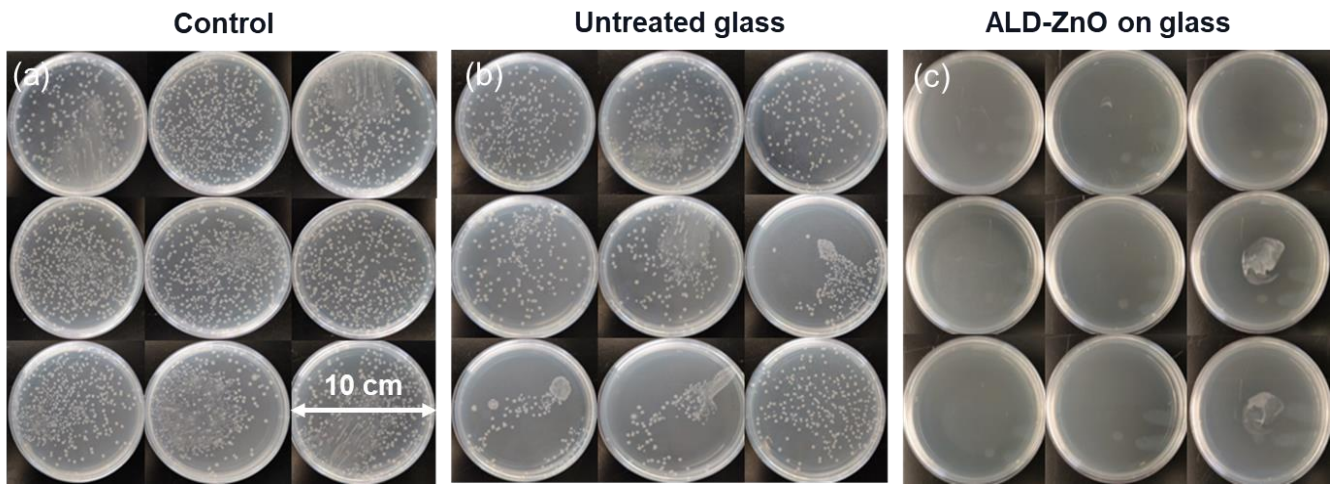

**Figure 6.** Optical images of cultured *E. coli* in the presence of (a) the control sample, (b) untreated glass, and (c) ALD-ZnO coated glass.

The antimicrobial performance of the ALD-ZnO sample in the light chamber was analyzed by counting bacteria colonies, which appear as white dots on agar plates (Figure 1). The control sample (containing only diluted bacteria) and the bare glass were also included in the evaluation to provide a baseline for the data analysis. Each sample has nine multiplication to ensure sufficient data quantity, and the results are compared in Figure 6. Upon visual inspection of Figure 6a-b, both the control and bare glass samples show apparent *E. coli* colonies growth. Moreover, the bare glass samples had uneven growth patterns in some of the Petri dishes. This uneven colony distribution is likely caused by water condensation inside the Petri dish, which washes away plated *E. coli* and creates an uneven distribution of bacteria colonies formation. However, based on the number of data points obtained from each sample set, these condensation-affected data can be compromised. The ALD-ZnO sample in Figure 5c demonstrates no visual growth of *E.coli* colonies, suggesting that the growth of *E.coli* were effectively limited.

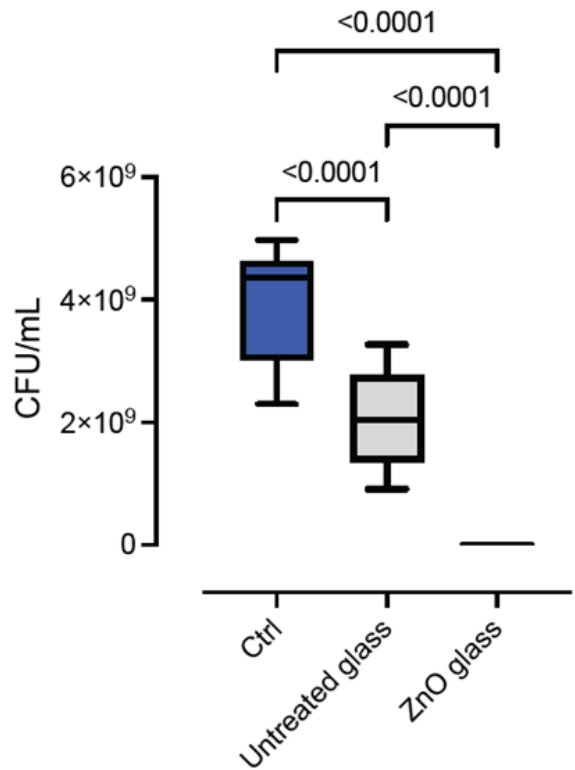

**Figure 7.** Count comparison between control (Ctrl), untreated glass, ALD-ZnO on glass, and combined box plots.

The data collected from Figure 6 are calculated into colony-forming units (CFU/ml), and the results are presented as a box plot in Figure 7. As indicated in the box plot, data of the control sample varies between $2.30 \times 10^9$ CFU/ml and $4.97 \times 10^9$ CFU/ml with a median of $4.36 \times 10^9$ CFU/ml. Data of the untreated glass span between $9.10 \times 10^8$ CFU/ml and $3.27 \times 10^9$ CFU/ml with a median of $2.04 \times 10^9$ CFU/ml. All ALD-ZnO data remain zero because no colony formed during the antimicrobial test. Figure 7 proves that the ALD-ZnO sample has completely prevented *E.coli* growth, which matches the behaviors shown in the photos (Figure 6). It also illustrates that the untreated glass samples pose a lower data span than the control sample, which suggests that the untreated glass potentially might have growth prevention of *E.coli*, even only to a small extent.

The obtained data are processed with statistic analysis techniques to validate further the confidence level of conclusions drawn from the box plot and data accuracy. The null-hypothesis significance testing, *P*-value test, with the alpha value set at 0.05, was performed to statistically verify the relationship between the control samples (Ctrl), untreated glass samples (untreated glass), and ALD-ZnO samples (ZnO glass). The calculated *P* value (indicated on the top of Figure 6) between untreated glass and control, ZnO glass and control, and untreated glass and control, are all smaller than 0.0001, which can be considered significantly smaller than the 0.05 alpha value. The incredibly small *P*-value for untreated glass and control samples indicates that the data of these two samples have a statistically significant difference. Considering that the untreated glass sample has a lower data span than the control samples, it is safe to conclude that the untreated glass has created constraints over *E.coli* growth.

Similarly, as zeros, the ALD-ZnO sample data also poses statistically significant differences between the untreated glass and the control sample. This not only proves that ALD-ZnO has an effective suppression on *E.coli* growth but also an indication that although untreated glass has limitations over *E.coli* growth, the effect is still significantly weaker than samples with ADL-ZnO coated samples. Bacteria reduction for untreated glass and ALD-ZnO coated glass are calculated using validated data, with respect to the control sample, resulting in 46.5 % and 100 % *BR*, respectively. Based on the analyzed results, it is safe to conclude that the prevention of *E.coli* growth in the ALD-ZnO sample is mainly caused by the existence and effects of ALD-ZnO coating instead of other causes.

In addition, the antimicrobial performance of commercial ZnO powders with a concentration of 4g/L in nutrient broth was also tested to provide a standard performance benchmark. The result is shown in Figures SI-4 and Figure SI-5. A comparison of the ALD-ZnO with the commercial ZnO powders indicated that the ALD-ZnO sample could reach an antimicrobial performance similar to the commercial ZnO powders. It suggests that the ALD-ZnO coating, even with a thin film thickness ($312.8 \pm 41.3$ nm), possesses strong antimicrobial performance as bulk ZnO powders. Moreover, we successfully applied

ALD-ZnO coating on cloth fabric with a 3D interconnected fiber structure (Figure SI-6), and EDX analysis discloses the uniform deposition of ZnO layers. This demonstration proves the feasibility of directly applying ALD-ZnO coating on temperature-sensitive personal protective equipment, suggesting the great potential of the ALD process for antimicrobial coating applications.

## 4. Conclusions

ZnO coating was successfully deposited on the glass substrate by using the ALD process of DEZ/$H_2$O at 150 °C. The ALD-ZnO thin film consisted of densely packed nano-plates at the microstructure level and was uniform across the substrate macroscopically. The crystal structure and chemical composition of the ALD-ZnO thin film were confirmed by XRD, Raman, and XPS analysis. Antimicrobial property evaluation revealed that, despite the thin thickness (~ 312.8 nm), the ALD-ZnO effectively prevented the growth of *E. coli* and possessed a similar antimicrobial performance as commercial ZnO powders. The main reason could be that, upon exposure to light, the ALD-ZnO could release reactive oxygen species to eliminate *E. coli* bacteria. The ALD-ZnO is promising for usage as an antimicrobial coating for various personal protective equipment, such as reusable masks.

**Acknowledgement**

This work was supported by the Nature Sciences and Engineering Research Council of Canada (NSERC), Canada Foundation for Innovation (CFI), BC Knowledge Development Fund (BCKDF), Alliance COVID-19 (ALLRP 554327-20), Peter Wall Institute for Advanced Studies (COVID-19 Wall Solutions Initiative 2020), Interior University Research Coalition (IURC), Myant Inc. and the University of British Columbia (UBC). The authors thank Dr. Sudip Shrestha at the UBC Fipke Laboratory for Trace Element Research (FiLTER) for supporting SEM and EDS analysis.

**Supporting Information (SI)**

**Antibacterial Activity of Zinc Oxide Thin Films by Atomic Layer Deposition for Personal Protective Equipment Applications**

Li Tao, Kinga Vojnits, Man In Lam, Xuejun Lu, Sepideh Pakpour,[*] Jian Liu [a,*]

*[a] School of Engineering, Faculty of Applied Science, The University of British Columbia, 3333 University Way, Kelowna, BC V1V 1V7, Canada*

Corresponding authors' emails: sepideh.pakpour@ubc.ca (S. Pakpour), jian.liu@ubc.ca (J. Liu)

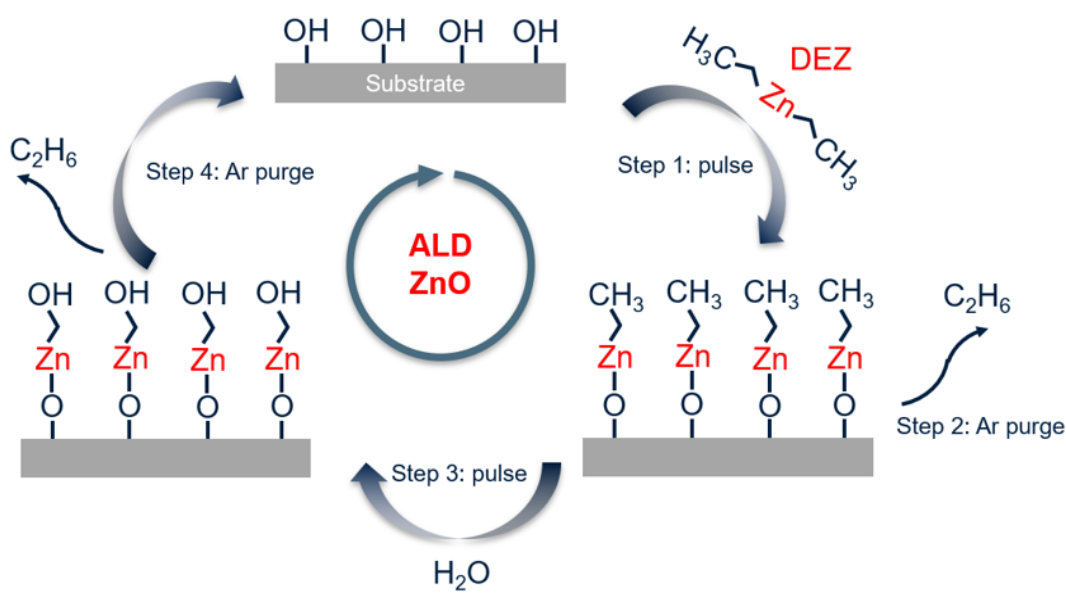

**Figure SI-1**. One ALD cycle of ZnO on the glass substrate by alternatively introducing diethylzinc (DEZ) and water ($H_2O$).

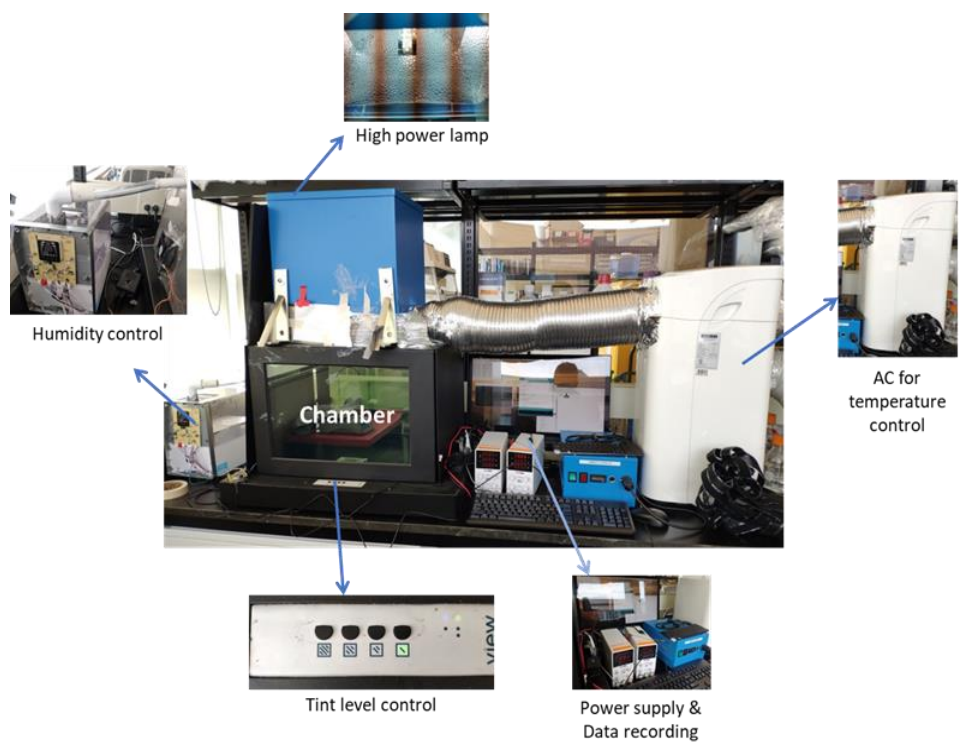

**Figure SI-2.** Artificial sunlight chamber with humidity, temperature, light intensity control, monitor, and recording system.

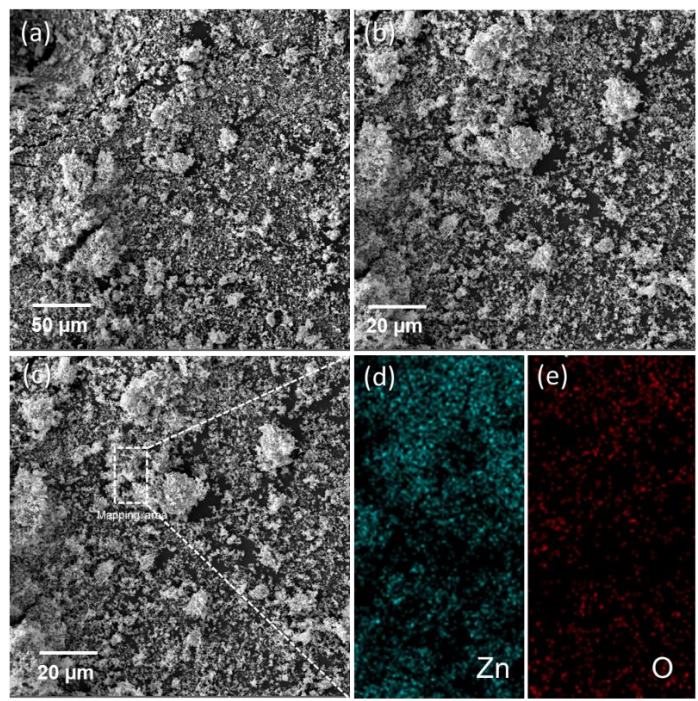

**Figure SI-3**. (a-c) SEM images and EDX elemental mapping (d) Zn and (e) O elements of commercial ZnO powders (GRADE CR-4, Ever Zinc).

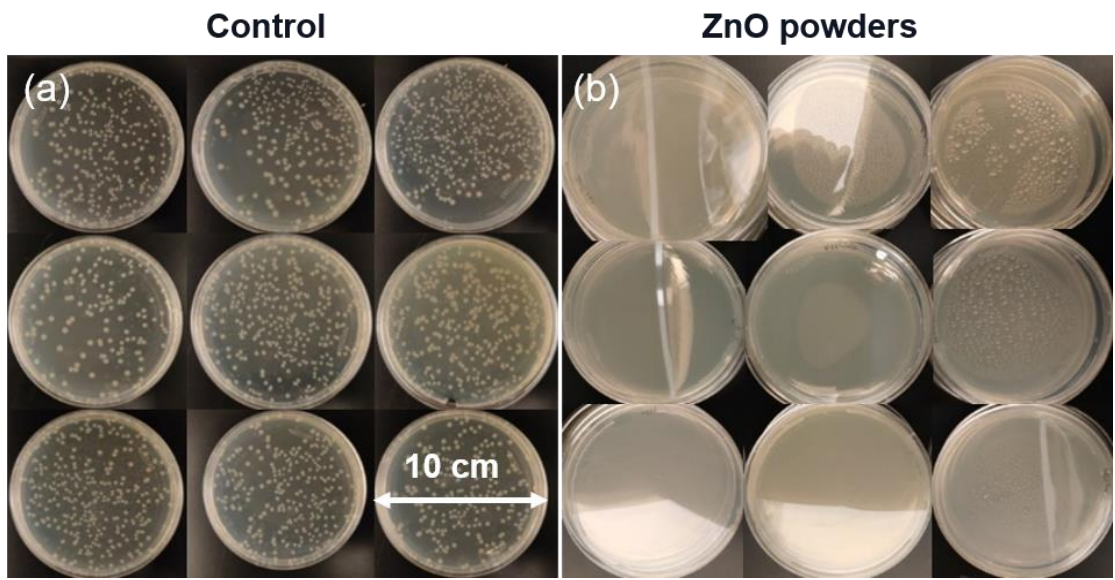

**Figure SI-4**. Optical images of cultured *E. coli* in the presence of (a) the control sample and (b) the commercial ZnO powders.

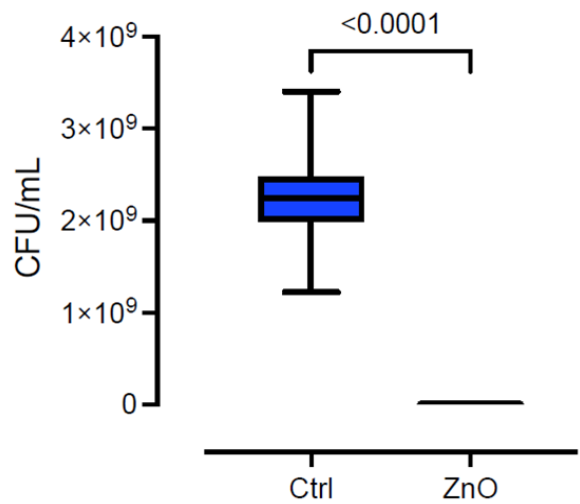

**Figure SI-5**. *E. coli* colonies counting test result of commercial ZnO powders.

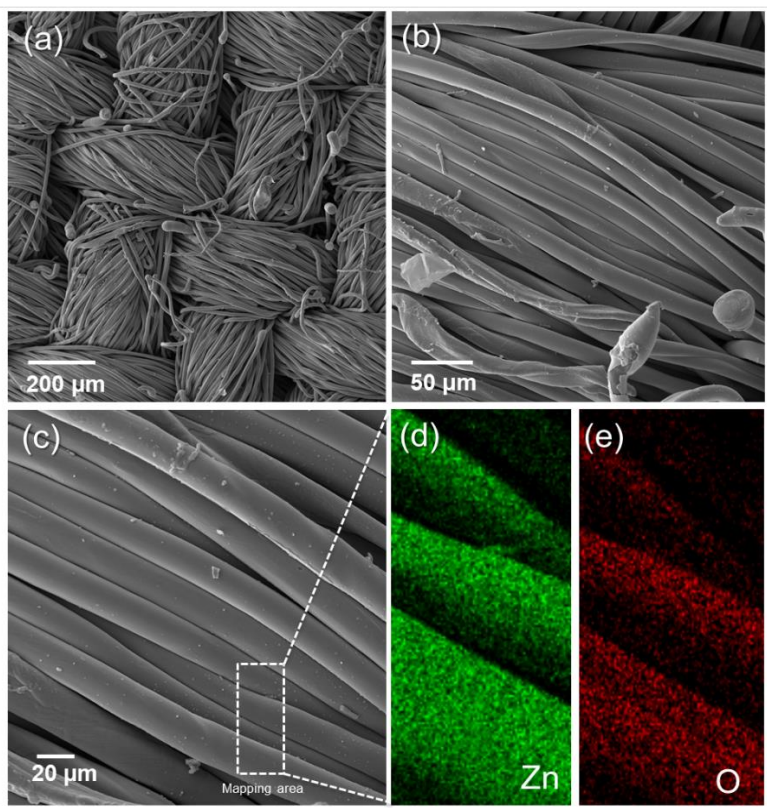

**Figure SI-6**. (a-c) SEM images and EDX elemental mapping (d) Zn and (e) O elements of ALD-ZnO on cloth fabric.

**Table SI-1**. Film thicknesses of ALD- ZnO on glass measured by a surface profilometer.

|         | ALD-ZnO film thickness on the glass substrates (nm) |
|---------|-----------------------------------------------------|
| Site A  | 331.8                                               |
| Site B  | 382.5                                               |
| Site C  | 261.8                                               |
| Site D  | 292.2                                               |
| Site E  | 296.12                                              |
| Average | 312.8 ± 41.3                                        |

# Reference

[1] S. Ferraris and S. Spriano, "Antibacterial titanium surfaces for medical implants," *Materials Science and Engineering: C* vol. 61, pp. 965-78, Apr 1 2016.

[2] H. M. Yadav, J.-S. Kim, and S. H. Pawar, "Developments in photocatalytic antibacterial activity of nano $TiO_2$: A review," *Korean Journal of Chemical Engineering,* vol. 33, no. 7, pp. 1989-1998, 2016.

[3] S. M. d. Amorim *et al.*, "Antifungal and Photocatalytic Activity of Smart Paint Containing Porous Microspheres of $TiO_2$," *Materials Research,* vol. 22, no. 6, 2019.

[4] G. J. Tortora, B. R. Funke, and C. L. J. U. P. E. Case, "Microbiology: An Introduction, Global Edition," 2015.

[5] J. Endrino *et al.*, "Antimicrobial properties of nanostructured $TiO_2$ plus Fe additive thin films synthesized by a cost-effective sol–gel process," vol. 3, no. 5, pp. 629-636, 2011.

[6] H. M. Xiong, "ZnO nanoparticles applied to bioimaging and drug delivery," *Advanced Materials,* vol. 25, no. 37, pp. 5329-5335, 2013.

[7] H. Li, Y. Zou, and J. Jiang, "Synthesis of Ag@CuO nanohybrids and their photo-enhanced bactericidal effect through concerted Ag ion release and reactive oxygen species generation," *Dalton Transactions,* vol. 49, no. 27, pp. 9274-9281, Jul 21 2020, doi: 10.1039/d0dt01816c.

[8] C. J. Chung, H. I. Lin, H. K. Tsou, Z. Y. Shi, and J. L. He, "An antimicrobial TiO2 coating for reducing hospital‐acquired infection," *Journal of Biomedical Materials Research Part B: Applied Biomaterials: An Official Journal of The Society for Biomaterials, The Japanese Society for Biomaterials, and The Australian Society for Biomaterials and the Korean Society for Biomaterials,* vol. 85, no. 1, pp. 220-224, 2008.

[9] V. Kumaravel *et al.*, "Antimicrobial $TiO_2$ nanocomposite coatings for surfaces, dental and orthopaedic implants," *Chemical Engineering Journal,* vol. 416, p. 129071, 2021.

[10] L. Dyshlyuk *et al.*, "Antimicrobial potential of ZnO, $TiO_2$ and $SiO_2$ nanoparticles in protecting building materials from biodegradation," *International Biodeterioration & Biodegradation,* vol. 146, p. 104821, 2020.

[11] K. S. Khashan, G. M. Sulaiman, and F. A. Abdulameer, "Synthesis and antibacterial activity of CuO nanoparticles suspension induced by laser ablation in liquid," *Arabian Journal for Science and Engineering,* vol. 41, no. 1, pp. 301-310, 2016.

[12] S. J. Jung *et al.*, "Bactericidal effect of calcium oxide (Scallop‐Shell Powder) against Pseudomonas aeruginosa biofilm on quail egg shell, stainless steel, plastic, and rubber," *Journal of Food Science,* vol. 82, no. 7, pp. 1682-1687, 2017.

[13] E. Torres Dominguez, P. H. Nguyen, H. K. Hunt, and A. Mustapha, "Antimicrobial coatings for food contact surfaces: Legal framework, mechanical properties, and potential applications," *Comprehensive Reviews in Food Science and Food Safety,* vol. 18, no. 6, pp. 1825-1858, 2019.

[14] Z. Cao *et al.*, "Citrate-modified maghemite enhanced binding of chitosan coating on cellulose porous membranes for potential application as wound dressing," *Carbohydrate Polymers,* vol. 166, pp. 320-328, 2017.

[15] P. Kelly *et al.*, "Comparison of the tribological and antimicrobial properties of CrN/Ag, ZrN/Ag, TiN/Ag, and TiN/Cu nanocomposite coatings," *Surface and Coatings Technology,* vol. 205, no. 5, pp. 1606-1610, 2010.

[16] P. Ramesh, J. Zhao, J. Provine, and M. Rincon, "Atomic Layer Deposition of Zinc Oxide," *Basic Solid State Physics*, vol. 257, no. 2, 2014.

[17] T. Suntola and J. Antson, "Method for producing compound thin films," US4058430A, 1977.

[18] J. Cai *et al.*, "A revisit to atomic layer deposition of zinc oxide using diethylzinc and water as precursors," *Journal of Materials Science,* vol. 54, no. 7, pp. 5236-5248, 2019.

[19] V. Miikkulainen, M. Leskelä, M. Ritala, and R. L. Puurunen, "Crystallinity of inorganic films grown by atomic layer deposition: Overview and general trends," *Journal of Applied Physics,* vol. 113, no. 2, p. 2, 2013.

[20] T. Tynell and M. Karppinen, "Atomic layer deposition of ZnO: a review," *Semiconductor Science and Technology,* vol. 29, no. 4, p. 043001, 2014.

[21] Y.-Q. Cao, S.-S. Wang, C. Liu, D. Wu, and A.-D. Li, "Atomic layer deposition of ZnO/TiO$_2$ nanolaminates as ultra-long life anode material for lithium-ion batteries," *Scientific Reports,* vol. 9, no. 1, pp. 1-9, 2019.

[22] V. Nefedov, Y. V. Salyn, G. Leonhardt, and R. Scheibe, "A comparison of different spectrometers and charge corrections used in X-ray photoelectron spectroscopy," *Journal of Electron Spectroscopy and Related Phenomena,* vol. 10, no. 2, pp. 121-124, 1977.

[23] E. Janocha and C. Pettenkofer, "ALD of ZnO using diethylzinc as metal-precursor and oxygen as oxidizing agent," *Applied Surface Science,* vol. 257, no. 23, pp. 10031-10035, 2011.

[24] K. P. Lawrence, T. Douki, R. P. Sarkany, S. Acker, B. Herzog, and A. R. Young, "The UV/visible radiation boundary region (385–405 nm) damages skin cells and induces "dark" cyclobutane pyrimidine dimers in human skin in vivo," *Scientific Reports,* vol. 8, no. 1, pp. 1-12, 2018.

[25] Y. Ibuki, Y. Komaki, G. Yang, and T. Toyooka, "Long-wavelength UVA enhances UVB-induced cell death in cultured keratinocytes: DSB formation and suppressed survival pathway," *Photochemical & Photobiological Sciences,* vol. 20, no. 5, pp. 639-652, 2021.

[26] Y. Jiang, L. Zhang, D. Wen, and Y. Ding, "Role of physical and chemical interactions in the antibacterial behavior of ZnO nanoparticles against E. coli," *Materials Science and Engineering: C,* vol. 69, pp. 1361-1366, 2016.

[27] L. Zhang *et al.*, "Mechanistic investigation into antibacterial behaviour of suspensions of ZnO nanoparticles against E. coli," *Journal of Nanoparticle Research,* vol. 12, no. 5, pp. 1625-1636, 2010.